# Case Studies: Business and Technical Perspectives in Migration of Legacy Systems to Service Oriented Architecture

**Maulahikmah Galinium**[1] and **Negar Shahbaz**[2], Non-members

**ABSTRACT**

In adoption process of Service Oriented Architecture (SOA), the legacy systems of a company can not be neglected. The reason is the legacy systems have been deployed in the past and have been running critical business processes within an enterprise in its current IT architecture. However not all migration process of legacy systems to SOA has been successfull. Highlighting the right factors to reach legacy systems migration success in a specific company is the key value. The main adopted research method in this study has been interviewed for different companies with different enterprises including bank, furniture, engineering and airline companies in Europe. Through separate interviews, critical success factors of migrating legacy systems into SOA have been collected and identified in each case company. Finally collected results are analyzed and presented as the recognized factors affecting successful migration of legacy assets into SOA from business and technical perspectives.

## 1. INTRODUCTION

Service Oriented Architecture (SOA) is one of the current technologies in the field of information system architecture. Reference [1] described SOA as a system that is divided into several sub-systems based on the functionality in the business process of the company. Then all the functionalities can be used as an interoperable service in order to run the business process. Like other technologies, SOA causes pro and contra eventought the use of SOA offers efficiency, reusability, agility and productivity of an enterprise [2].

The SOA adoption process usually is not an easy task because it will revolutionize the company and its information systems, which are the core of the company. Therefore both IT technicians and the business executives must follow the whole process of adoption from beginning until post-adoption phases to study whether the SOA is succesfully adopted or not in the company's information system [3, 4].

This research focuses on SOA adoption from the legacy system. It refers to the existing information system that has been developed in the past and running the significant role of the business process in the company [5]. It is backbone and core of the organization's information flow [6].Therefore to migrate the legacy system into SOA, we can not neglect the importance of legacy system.

The success factors of the migration process will be discussed in this research. In general, the success factors can be seen from two different perspectives. The first perspective is an IT or technical perspective. It will be more focused in the technical issues of the migration process from the legacy information systems of the company into the SOA framework architecture. The second perspective is from the business point of view. It will study the business internal and external factors of the SOA adoption process whether the adoption still aligns with the primary business goals of the company or not.

The following sections describe how the research is conducted and analyzing the results of all possible success factors in the migration legacy system into SOA by sharing experience from five different case companies in different business area.

## 2. RELATED WORK

Reference [7] mentioned the potential to reuse the legacy systems as components in SOA by exposing the legacy system's functionality as services. U.S. Department of Defense (D.O.D) has adopted the factor of reusing the legacy system component [7]. To reuse the legacy systems, evaluating legacy assets [8] is important in the context of migration to SOA. Furthermore some researches have discussed the migration strategy of legacy systems into SOA such as black box [9], wrapping methodology [10], reengineering approach [11] and multi-tier architecture [12] strategies.

The business process of the company, which is run by the legacy system, is also need to be considered [13]. It should not be changed drastically when adopting SOA because the business process has significant business value for the company. IT infrastructure of a company play an important role in the success of its business process design, implementa-tion and redesign [14]. Reference [15] studied that SOA governance is also essential factors to succeed in the SOA adoption. It includes leadership, funding, ownership, policies, guidance for SOA infrastructure providers and application developers.

Many companies have common problem in SOA adoption because they start the SOA adoption process based on the IT perspective instead of a business one. The implementation might appear successful at that time but the impact after adopting the new architecture could be not aligned with the business goals of the company [3]. Therefore strong project planning in both technical and business perspectives is required because such mistake could cause the growing cost of IT system without any return on investment for the company. Reference [16] proposed SOA adoption roadmap management to have better planning in SOA implementation.

Reference [17] has mentioned that when it comes to migration to SOA, no one can ignore the important need for careful budgeting. SOA migration can cost huge amounts of funds and can take long periods to complete but it can turn out worth investing on through appropriate funding. Reference [18] has mentioned that SOA budget planning considering the mentioned dimensions to SOA adoption can seriously contribute to the success of adopting and implementing a service-oriented architecture. On the other hand, lack of insight and foresight on the required migration budget can push enterprises towards spending (better to say waste) their money on aspects and areas which are not the right fields to guarantee a long-term success for their SOA.

Considering the literature review and the previous research carried out in this field, we have come to the following factors as the possible factors affecting a successful migration of legacy systems into SOA:

• SOA Migration Strategy: Before conducting migration, it is necessary to consider the migration strategy either being technical or business-oriented in which companies decide whether to reuse the legacy assets as components or web services or to apply redevelopment.
• Potential of Legacy Systems to be integrated with Service-Oriented Architecture: This is also one technical factor required to be taken into consideration before starting the migration. This is due to the fact that not all the legacy applications have the potential to be integrated and reused as components or services in SOA.
• SOA Governance: SOA Governance can define boundaries and regulations for legacy migration through agreements known as SLAs. SLAs can be technical or business-oriented.
• Business Process of the Company: Business Process of a company defines the way its routine tasks are run and it is usually handled by the legacy applications in a company. It can be seen as a business-oriented factor since it is directly related to the way business is run.
• Budget Plan of SOA migration: Budgeting and Resources also play an important role in our findings since migration is a costly process in terms of monetary, human and time resources. This factor lays on the strategic and business-oriented decision making of the company.

The findings above are verified in two perspectives, the business perspective and the technical one. In this research, the factors have been used to design the interview questions and have been under study in the five cases of this study.

## 3. METHODOLOGY

The research method will be presented and motivated in detail in this section regarding how the data has been collected and analyzed. It is generally presented in Fig. 1. In this research, we aimed to interview the company's business and technical members who had been involved in the SOA adoption project's leading team in order to have both business and technical perspectives on the real-life factors leading migration of legacy system into SOA regardless of the final outcome having been failure or success. The interview results are our primary data while supporting documentations from the companies as our secondary data are also required to have better knowledge about their SOA adoption process. Interview method is selected because we conducted qualitative research that is used to understand the experience of the companies members in SOA adoption process.

We adopted the seven stages of research interview process (see Fig. 2) including thematizing, designing, interviewing, transcribing, analyzing, verifying and reporting [19]. It starts from choosing the interview topic based on the research goals. Then in the second stage, the overall planning and preparation to conduct the interviews are prepared including planning for the questions, how to transcribe, analyze and verify the result and how to write the report. When preparing the interview questions, we considered both technical and business-oriented themes. The questions about potential of legacy system, its architecture, quality of service, technical risks and maintenance, governance and the strategy of migration are considered in the tecnical theme while we had some questions about the company's business process and budgeting plan in the business theme. We also asked about the general monitoring and testing of the SOA adoption trend as well as the challenges which could come when carrying out the migration of legacy system to SOA.

***Fig.1:***  *Research Methodology*

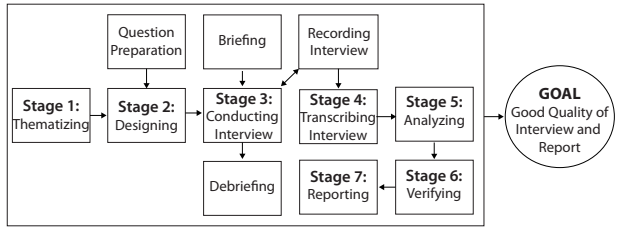

***Fig.2:*** *Interview Process*

Furthermore in the third stage during the interview process, briefing and debriefing are important to make the interviewee understand the research study and its purpose and also to get the feedback from the interviewees. In the fourth stage, the interview results are transcribed into a written script. We transcribed the interview individually then we discussed the result of the transcribing. It is used to maintain the reliability of the interview result and minimize the possible error. In the fifth stage, analyzing the evidence from the companies is required. In this research's analyzing, theoretical propositions strategy [20] is used due to the fact that some factors already existed in SOA adoption process but they are not explicitly mentioned as success factors in the literature. Therefore we carried out a literature review to get some clue about the factors then we investigated whether these factors fitted to the companies or not. In this research, other specific success factors are also discovered in each company's empirical findings. We used cross case analysis [20] as our analytical technique by aggregating our findings across a series of individual studies. Using cross case analysis, we used word tables [20] that display the collected data from the companies based on our uniform framework.

Research validity and reliability also need to be considered in the sixth stage of this project. There are four criteria which are related to the quality of a case study, including construct validity, internal validity, external validity and reliability [20]. For construct validity, we analyzed the result of each study based on the theory of migration of legacy sytem into SOA that we found in the literature. In this case, internal validity is based on each study while for external validity, we established the domain to which a study's findings can be generalized. We also validated our findings analysis by checking it with the interviewees whether they agree with the analysis or they want to add some more information. We also measured the reliability of our interview transcriptions by comparing two version of transcriptions which were made by us. Finally, in the seventh stage, we made a report of the result of this research.

## 4. EMPIRICAL FINDINGS IN FIVE CASE COMPANIES

This section discusses the empirical findings as the result of conducted interviews in the five different companies including the analysis and discussion for each case company. These five different companies consist of 2 banks, 1 airlines company, 1 furniture company and 1 manufacturing company. We did our research in different type of companies in order to know different experience in different areas. All companies have quite big business size because each of them has more than 1000 employees and has company branches abroad.

### 4.1 European Bank

The bank started its SOA experience by adopting a proprietary SOA with the bank's own protocols and then redefining the infrastructure and starting with SOA development in 2002. The bank have got some benefits in SOA adoption including flexibility, faster time to market and reusability of the services. The bank started SOA adoption by creating a new banking system based on completely new-developed code but in the beginning of the project, the bank wrapped all services of the legacy system in which the other applications could access. In the next phase, they gradually changed the legacy code into the new-developed code, which is called a refactoring approach. In this approach, there are massive data transformation effort from legacy system formats to new format. This problem was a challenge for the bank because the XML transformation were costly to achieve the required performance, stability and scalability. In order to conduct the SOA adoption process, the bank considers the following success factors:

1. *Strategy of migration to SOA*: The interviewee mentioned that the strategy is the most important factor in migration of their legacy system to SOA. The company used the refactoring approach.

2. *Business Process of the Company*: The initial business process of the bank was not appropriate anymore, neither in terms of time to market nor in terms of supporting the new business model. Since the bank would turn from a local bank in a country into a global bank, the initial business process and its legacy IT system did not fit together anymore.

3. *Potential of Legacy System*: Good application structure of legacy system in the bank also contributed to success in SOA adoption. Level of documentation and code quality of the legacy system were also needed to make a reliable plan for the migration especially when the original developer is not anymore involved in the project. In addition, size and complexity, support software required, reusability factors, scale of required changes in legacy system to SOA, service abstraction and service discoverability were also important characteristics of legacy system in SOA adoption.

4. *Legacy Architecture*: Security and architecture of the legacy system can affect the difficulty of the migration effort. If the legacy applications have been well-structured and well-defined, they are expected to have good pattern which will ease the migration.

5. *Close Monitoring*: Closely monitoring the project progress and outcome is also important, for example the bank realized that the complexity of the relationship between services and their sub-services were not managed very well due to not enough focus on the service orchestration.

6. *SOA Governance*: Business and application architecture must describe the success criteria or business value of the SOA. The interviewee mentioned that IT must also focus on non-functional requirements (NFR), such as security, auditability, authenticity of messages, performance and throughput. Beside that, the company also considered a Service Level Agreement (SLA) in which the following aspects were considered: availability, throughput, performance and all the interdependencies of all components.

7. *Budgeting and Resources*: The interviewee mentioned that they had underestimated the required budget because the initial budget including the service wrapping costs was increasing due to the subsequent replacement of the legacy code behind the service layer.

8. *Dependence on Commercial Products*: This factor can make the integration more difficult.

**4. 2  Airlines Company**

The company started to do SOA adoption in the late 1990's and started to develop web services in 2003. In this company, SOA adoption means adding new functionality when implementing web services integrated with the existing legacy systems. They have reused the legacy systems with additional new functions which are integrated with the existing systems. The main purpose of SOA adoption in the company have been based on market purposes as well as feasibility and reusability. They decided to use web services to implement SOA, which is used to implement the company's operational data stores and generic services. Using web services had brought more flexibility to the communication between the systems and within the entire environment. In order to conduct the SOA adoption process, the company considers the following success factors:

1. *Strategy of migration to SOA*: The role of migration strategy in the company is building up a reusable SOA. The company has made good effort in having their legacy applications reused in various projects.

2. *SOA Governance*: SOA governance is important to control the growing of the SOA. The company has defined their own SLAs to avoid losing control and ending up with a spaghetti dish of web services in both their implementation and architecture.

3. *Potential of legacy systems*: The company believes that reusability factor of the legacy systems is the most important role when integrating them with SOA. Scale of required changed for integrating the legacy code with web services and level of documentation also have important roles and can ease the integration purpose at the business and technical levels. The size, complexity and code quality of the legacy applications as well as the supportive software required for SOA adoption are also influencing factors in the legacy system integration with SOA.

4. *Close Monitoring*: The company does the monitoring of their progress' project through architecture reviews made in every phase of each project.

**4. 3  Swedish Company**

The company started adopting SOA technology in around 2004. Using their initial business cases, the company came to initial estimation about cost, duration and benefits of SOA adoption in the company. However none of them have been fulfilled successfully as they were predicted. Almost none of the business services have been reused. Their initial SOA infrastructure has been only for the company's internal service and use. One of the first services created in their SOA was a new store system. Eventhough the new system has been very successful, they still use the old store system in the back-end. Thus the SOA adoption in the company was not successful in terms of reusability and budgeting. However the SOA adoption have been able to deliver the functionality needed by the company. The systems also have a stable and centralized environment with more controllable systems to support the company's growth. The following factors affects migration of their legacy systems to SOA in the company:

1. *Business process of the company*: The company has a culture in which technology is never allowed to control the business process. Therefore they have to create their own IT systems that follow their business process.

2. *Potential of the legacy systems*: The company needs maintainability skills in order to find resources in the legacy code. The company divide the potential of legacy system into two perspectives, which are the business level and the technical level. For technical level, they consider scale of changes required, support software required, size and complexity, code quality and level of documentation of their legacy systems. Reusability factors, service abstraction and service discoverability are important in the business level.

3. *Legacy Architecture*: The company planned to transform the legacy systems into different thinking and divide them into services that could be written in a reusable way. In this case the architecture of setting up the legacy system is important aspect because the logic and data were together and then the company want to separate them.

4. *Strategy of migration*: The main SOA adoption

strategy in the company was redeveloping. They had to rebuild the system by using the structure of the legacy system, part of its functionality or the flow of its code regardless the language they were developed in.

5. *Dependence on commercial products*: Commercial product had an affect on the migration but it was not very significant for the company because the company has developed almost every system by themselves.

6. *SOA Governance*: For this company, SOA governance has been used to manage all resources in the project in order to ensure service quality, consistency, predictability and performance.

7. *Testing*: Testing is important to validate and verify the quality of the services in terms of performance, reliability and security. The company need to schedule and maintain their test environments so that each component is in the correct direction when being used.

8. *Budgeting and Resources*: SOA adoption has been a costly project in the company in terms of finance, time, personnel and hardware. In this case, the real cost and time were around 500% more than the initial estimated budget.

### 4.4 Manufacturing Company

The company is a manufacturing company which has representations in 130 countries around the world. It is one of the early adopters of the web service technologies in 1988. The company has technically succeeded in migration of its legacy systems to SOA. They deployed SOA by reusing their legacy systems and reducing redundancy. The company has almost all existing environments in SOA due to mergers and acquisitions (M&A) and an early adoption of IT system with local presence. In terms of SOA adoption success, they said that the agility part is critical. It is caused by integration which required a standardized information architecture. The following factors affects migration of their legacy systems to SOA in the company:

1. *Technically skilled personnel*: The skilled personnel played an important role in the success of SOA adoption. This has been due to the fact that the standardization is driven from IT.

2. *Information Architecture*: Low maturity of the information architecture of the company can lead to failure because SOA requires a global view of the information object. Information architecture is required to be handled in a structured way from the business side according to the requirements from processes and capabilities.

3. *Business Process*: The combination of business process and information architecture have also been very important in this company. Their SOA approach seldom has problems on the technical level and they believes that the migration process success is 70% dependant on the business side.

4. *SOA Governance*: This factor is about governance over information and services. They have established SLAs for each service in different solutions, for example to cover information delivered from a legacy system or to separately cover the integration phase.

5. *Potential of Legacy systems*: The company has used an approach to combine the components from legacy systems with new components from scratch when adopting SOA. Since the business logic residing in its legacy systems, the logic had to be used as much as possible.

6. *Legacy Architecture*: Their older architecture was founded in the 1970s based upon consumption of information in their systems and not scaled according to the current requirements. The requirement was to open up the business logic out of some of the legacy systems. There was a problem concerning agility. SOA will increase the complexity and the number of services required other than governance and coordination.

7. *Strategy of Migration*: the strategy of SOA adoption in the company based upon the business requirements which have a globally harmonized shared information. The strategy was focusing on standardizing the information architecture.

8. *Monitoring*: The company's IT team always monitor the SOA adoption results centrally for tracing and tracking purposes.

### 4.5 UK Bank

This company started adopting SOA in 2002. By using SOA, the company expected to obtain some benefits including efficiency, transformation, efficiency in terms of speed to deliver business services and ease of creating business process. The company considered the legacy system integration into their new components. Therefore the company needed people who had business analysis skills as well as in system integration. Furthermore, the company used Brownfield [5] in developing new systems in SOA as the immediate presence of legacy system. It can utilize an existing environment and follow the existing patterns in order to minimize risk. It is also used to analyze the application and the data in the system in order to know what and how to transform the legacy system. The following factors are their success factors of migrating legacy systems into SOA:

1. *Business process*: It is the most important factors in the company because SOA adoption required changes in their business process. The way SOA works is by separating service specification from the implementation to help their business.

2. *Legacy architecture*: The business process of the company is dependent on the legacy architecture. They need to divide the business process into two groups of the ones which can be reused and the ones which require new services.

3. *Budgeting and resources*: The company need to

analyze the cost estimation and number of functions required. It is required to understand the size and complexity of the legacy systems in order to know what to be migrated and how big the migration is.

4. *Potential of legacy systems*: Size and complexity are the most important feacure of the legacy systems for the company. They also need to know what potential services are hidden from the legacy system and what kinds of services are needed to be exposed and then how to concert it into new services.

5. *Strategy of migration*: The main strategy is to extend the services. They look at the new services which are defined by the business process of the company. They create the centre of the system and then the new services would come around that.

6. *SOA Governance*: They started creating SLAs when starting to migrate their legacy systems into SOA.

7. *Monitoring*: The company monitor the migration process all the time in order to help the strategy of migration to move forward with the SOA. Early signs of failure were also noticed.

## 5. RESULTS - SUCCESS FACTOR MODELS

In this section, we describe our results as a success factors model. We have used the theoretical propositions strategy. We carried out a literature review to learn about the possible factors which can influence conducting the migration of legacy systems into SOA. In our propositions, the affecting factors include potential of legacy systems for migration, strategy of migration, SOA governance, business process and budgeting. We then used interviews and their empirical data to investigate if those played a role in the companies. To analyze our empirical findings, we did a cross case analysis [20] using word tables and came to some results. We categorized the success factors into three categories including technical, business and both technical and business perspectives.

Table 1 shows the factors which are directly related to Information System technology when the companies migrate their legacy systems. We discovered seven factors affecting migration of legacy system in the case companies. Those factors include the potential of legacy systems for migration into SOA, legacy system architecture, information architecture, dependence of the legacy assets on commercial products, close monitoring, testing and technical skills. The reason for this categorization has been the direct relevance discovered based on the findings of the interviews between the mentioned factors and the technical aspects of migration rather than the business-oriented aspect [5].

Table 2 shows the business-oriented aspect of migrating legacy assets to SOA. There are two success factors in this aspect including business process and budgeting and resources of the company. Business process represents the routine of the company for handling the tasks of the company business while budgeting and resources are used to support the company throughout the SOA implementation. Furthermore, we discovered two other affecting factors, strategy of migration and SOA governance that could be categorized as both technical and business-oriented aspects. Strategy of migration plays an important role in selecting a proper technical strategy for adopting SOA beside the role it plays in the business decisions related to the migration. SOA governance can also be relevant to both perspectives through defining both business-oriented and technical SLAs. Table 3 displays the role of these factors in the five case studies.

Besides the success factors, during the interview, we also discovered eight characteristics of the legacy systems which can affect their migration into SOA. These factors include the size and complexity of the legacy applications, their level of documentation, scale of changes required, support software required, reusability factors, service abstraction, service discoverability and code quality. Table 4 shows the priority ranking of the characteristics for each case company from 1 (most important) to 8 (least important). We can see that there are three characteristics, which have always been mentioned in the top five characteristics in all of the case studies. These sub factors include size and complexity, reusability factor and level of documentation. Other sub factors such as scale of changes required, support software required, service abstraction, service discoverability and code quality have also been important.

Fig. 3 illustrates the success factors according to our cross-case analysis which has been based on our empirical data findings as well as our literature review. *Business Process* of a company defines the way its routine tasks are run and it is usually handled by the legacy applications in a company. Therefore it is important to consider the company's business process when it comes to migrating the legacy assets into SOA. *Budgeting and Resources* also play important roles because migration is a costly process in terms of monetary, human and time resources and therefore it is important for the companies to consider their budget and resources boundaries to support SOA adoption.

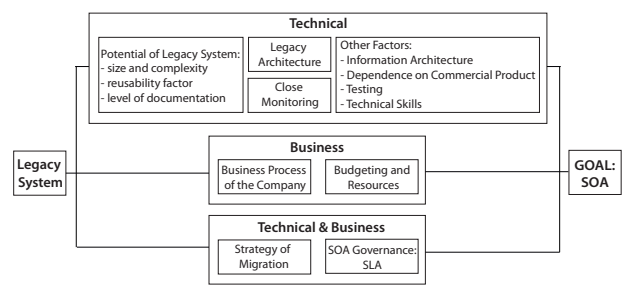

***Fig.3:*** *Success Factors Model*

**Table 1:** *Success Factors in Technology Perspective*

| Company | Technical Perspective Factors | | | | | | |
|---|---|---|---|---|---|---|---|
| | **Potential of the Legacy Systems** | **Legacy Architecture** | **Information Architecture** | **Dependence of the Legacy on Commercial Products** | **Close Monitoring** | **Testing** | **Technical Skills** |
| **European Bank** | It is their second factor. Good application structure of more than 25 years-old legacy systems contributed to success in migration of legacy systems to SOA. | This factor is considered as a part of the potential of their legacy systems. The architecture could affects the difficulty of the mgiration effort. | this factor is not mentioned. | This factor can make integration more difficult. | They consider it as a factor especially in monitoring the project progress and outcome. | This factor is not mentioned. | This factor is not mentioned. |
| **Airlines Company** | This is the most important factor due to the reusability. There are some characteristics in the legacy systems which can play important roles in their potential and the way they can be integrated with SOA. | It is not considered as a factor. | This factor is not mentioned. | It is not considered as a factor. | Monitoring of the progress projects was done through architecture reviews mande in every phase of each project. | This factor is not mentioned. | This factor is not mentioned. |
| **Swedish Company** | The migrations is always seen in the context of how the legacy system can be used within SOA. | This factor is important for them to learn whether the logic and data were together in the legacy artchitecture or not. | This factor is not mentioned. | This is not a significant factor. The company could deal with oracle licenses. | It is considered as a factor. | Only this company considered it as a factor. They need a testing environment when they want to integrate with the other system. | This factor is not mentioned. |
| **Manufacturing Company** | They combined components in legacy systems and new components during SOA adoption. | The older architecture in the company was founded based on consumption of information in their systems and not scaled according to the current requirements. | They mentioned that SOA adoption will not be success until the information architecture is handled in structured way from the business. | It is not mentioned as a factor. | The company could always monitor result centrally for tracing and tracking. | It is not mentioned as a factor. | Only this company mentioned this factor because the standardization of their system was driven from IT. |
| **UK Bank** | They needed to know how much the legacy system constrained the migration process. | The business process of the company depends on the legacy architecture. | This factor is not mentioned. | This is not considered as a factor. | The bank involves monitoring mode all the time in order to help the strategy of migration in the company to move forward with the new system. | This factor is not mentioned. | This factor is not mentioned. |

*Potential of the legacy applications* is required to be taken into consideration before starting the migration. This is due to the fact that not all the legacy applications have the potential to be integrated and reused as components or services in SOA. Therefore, there are some sub factors to consider when it comes to the potential of legacy systems for integration in SOA, including size and complexity, reusability factor and level of documentation, scale of changes required, support software required, service abstraction, service discoverability and code quality. *The architecture of the legacy systems* [21] is also important when it comes to their integration with SOA because the business process of the company is dependant on the architecture of the legacy applications. Beside that, both logic and data of the legacy applications reside in the legacy architecture. *Close Monitoring* is another factor to consider when migrating the legacy assets. Monitoring the migration process can help notice early signs of failure and then take measures for fixing the existing errors. *Strategy of Migration* is one factor which is necessary to consider when integrating the legacy with SOA. It is based on the migration strategy either being technical or business-oriented that companies decide whether to reuse the legacy assets as components or web services or to apply redevelopment. *SOA Governance* has also been a factor in the migration process. It can define boundaries and regulations for legacy migration through agreements known as SLAs which can be technical or business-oriented.

Comparing the results of related works with our findings, we can see that they are not much different. There are some additional factors which were not mentioned in the related works. These other factors which include Information Architecture, Dependence on Commercial Products, Legacy Architecture, Close Monitoring, Testing and Technical Skills have also been important in the migration of the legacy systems in the five studies. Information architecture defines the flow of information in a company. Dependence of the legacy assets on commercial products can define some constraints when it comes to the integration of the legacy applications with SOA. The legacy architecture of the company is based on the information architecture and it also defines the business process of the company. Close monitoring for the migration process can be an affecting factor in the migration since it can help the migration board recognize sings of early failure and take the required measures to prevent it. Testing has been mentioned as a factor in one of the cases since the legacy components require a test environment before getting integrated in the SOA. Technical skills of the project team members can also be a factor in the speed and quality of migration.

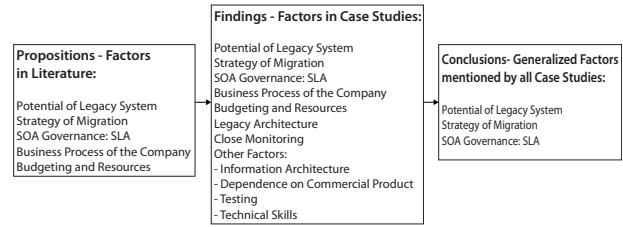

***Fig.4:*** *Success Factors Model*

## 6. CONCLUSION AND OUTLOOK

As our conclusion in this research study, we generalized the success factors found in our studies based on the common factors mentioned in all five case companies as well as the literature review. It must be mentioned that this generalization is still in the context of our research study. As for the specifications of this context, the selected companies had experienced a SOA adoption in which the legacy systems has been migrated into SOA. We prepared the theory for generalizations in the sense of collecting specific observations about success factors as basis for discovering similarities. We used cross case analysis technique to find these similarities. Among all factors mentioned in section 5. we discovered that only three of them have been applied and mentioned as affecting by all of the five companies (see Fig. 4) and therefore can be generalized as the affecting success factors when migrating legacy systems into SOA considering the context of transferability in this research. These factors include: *potential of legacy systems to be migrated into SOA*, *strategy of migration* and *SOA governance*.

For future research, it would be interesting to test the success factors identified in this research by performing quantitative research. Another interesting area would be comparing the two perspectives of IT and business people about the affecting success factors in migration of legacy systems to SOA. Such research can be either focused on one specific company or in general among some different companies. There is also still room for further exploration of our findings on how far they affect the migration process.

*Table 2: Success Factors in the Business Perspective*

| Company | Business Perspective Factors | |
|---|---|---|
| | **Business Process of the Company** | **Budgeting and Resources** |
| **European Bank** | It is one of the factors. The driver for creating a new core banking system was basically the fact that the business process was not acceptable anymore both in terms of time to market and supporting the new business model | It is one of the factors. If the legacy systems are not well maintained and not well understood anymore, it will be over budget. However everything have been on target regarding the budget plan in the company |
| **Airlines Company** | It is not a factor in the company | It is also not considered as a success factor because each project has its own separate budgeting and the estimation of costs was handled by the vendors for each project. |
| **Swedish Company** | It is the second most important factor considered by the company. The company has a culture that technology is never allowed to steer the business process | It is also a significant factor because the project cost a lot of hardware and human resources |
| **Manufacturing Company** | Combination of business process and information architecture made the most important factors in migration to SOA | It is not considered as a factor |
| **UK Bank** | It is the most important factor in the UK Bank. SOA adoption needed changes in the company's business process | Manage the cost is the second factor in the company. To be success, the company needed software analysis, including the cost estimation, number of functions needed, etc. The project's total cost consisted of hardware, software and service costs. Other costs could also from external consulting services and internal employee. |

*Table 3: Success Factors in both Business and Technology Perspectives*

| Company | Both Business and Technology Perspective Factors | |
|---|---|---|
| | **Strategy of Migration** | **SOA Governance** |
| **European Bank** | It is the most important factor. They used refactoring approach. | Business and application architecture define the success criteria and business value of SOA as well as the unifying and simplifying aspects of the various projects. |
| **Airlines Company** | The company highlighted the role of migration strategy in building up a reusable SOA. | This is the second important factor which is necessary in controlling the growing SOA in the company. They defined their own SLAs to avoid losing control and ending up with chaos web services in both the implementation and architecture |
| **Swedish Company** | It is the most important factor. They used redevelopment strategy. | This is one of the factors used to define SLAs to manage all the resources in the project in order to ensure service quality, consistency, predictability and performance. |
| **Manufacturing Company** | This factor is important especially in business strategy which is focusing on standardizing the information architecture to recognize the SOA prerequisites | In the company, this factor is about governance over information and services. |
| **UK Bank** | Their strategy is used to extend the services in the company. The IT team must especially be careful with their thinking and interpretation in creating new functions | They have SLAs for migrating the legacy systems into SOA. |

*Table 4: Characteristics of Legacy Systems*

| Company | Rangking of the sub factors in the Potential of Legacy Systems | | | | | | | |
|---|---|---|---|---|---|---|---|---|
| | Size & Complexity | Level of Documentation | Scale of Changes Required | Support Software Required | Reusability Factors | Service Abstraction | Service Discoverability | Code Quality |
| **European Bank** | same level | same level | same level | same level | same level | same level | same level | same level |
| **Airlines Company** | 4 | 3 | 2 | 6 | 1 | 7 | 7 | 5 |
| **Swedish Company** | 3 | 5 | 1 | 2 | Important in business | Important in business | Important in business | 4 |
| **Manufact. Company** | 2 | 4 | 3 | - | 2 | - | 5 | - |
| **UK Bank** | 1 | 4 | 6 | 7 | 5 | 3 | 2 | 8 |

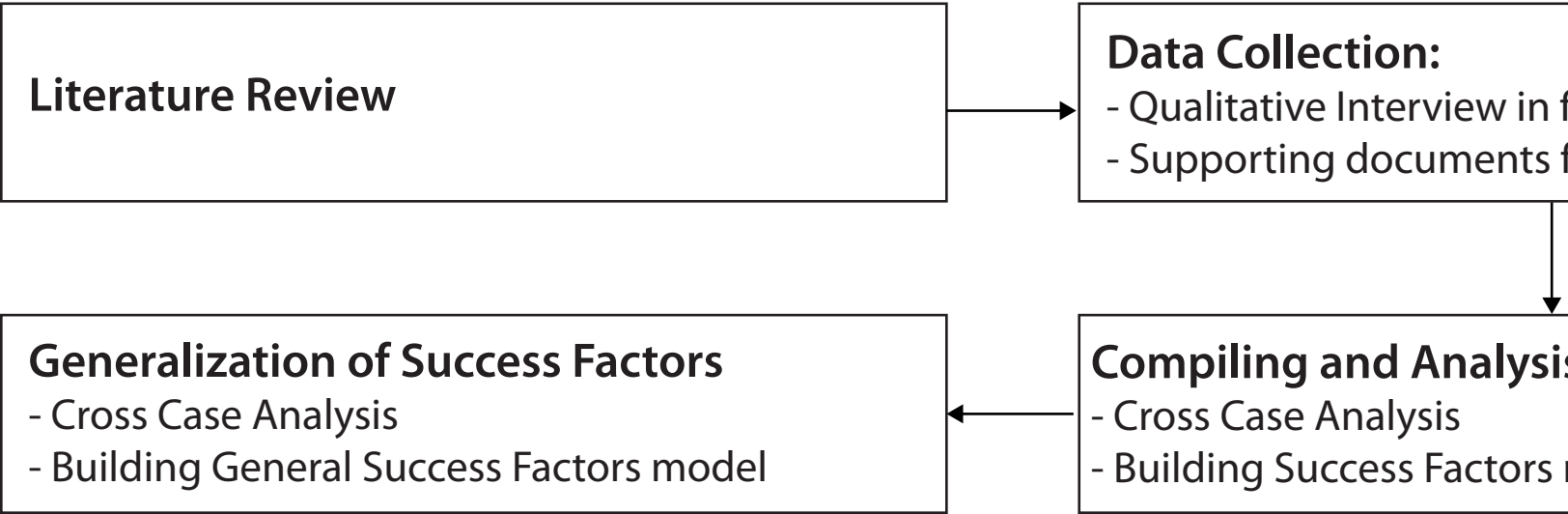